\documentclass[prb, preprint]{revtex4-1}
\usepackage{amsmath}
\usepackage{amssymb}

\usepackage{graphicx}   
\usepackage{tikz}
\usetikzlibrary{decorations.pathmorphing,patterns}
\usetikzlibrary{decorations.pathmorphing}
\usetikzlibrary{decorations.pathreplacing}
\usetikzlibrary{quotes,angles}
\begin{document}

\title{Conservation Laws and Energy Transformations in a Class of Common Physics Problems}
\author{Jonathan Bougie}
 \email{jbougie@luc.edu}   
 \affiliation{Loyola University Chicago, Department of Physics, Chicago,
IL 60660}
\author{Asim Gangopadhyaya}
 \email{agangop@luc.edu}   
 \affiliation{Loyola University Chicago, Department of Physics, Chicago,
IL 60660}
\date{July 10, 2019}

\begin{abstract}
We analyze a category of problems that is of interest in many physical situations, including those encountered in introductory physics classes: systems with two well-delineated parts that exchange energy, eventually reaching a shared equilibrium with a loss of mechanical or electrical energy. Such systems can be constrained by a constant of the system (e.g. mass, charge, momentum, or angular momentum) that uniquely determines the mechanical or electrical energy of the equilibrium state,  regardless of the dissipation mechanism. A representative example would be a perfectly inelastic collision between two objects in one dimension, for which momentum conservation requires that some of the initial kinetic energy is dissipated by conversion to thermal or other forms as the two objects reach a common final velocity. We discuss how this feature manifests in a suite of four well-known and disparate problems that all share a common mathematical formalism.  These examples, in which the energy dissipated during the process can be difficult to solve directly from dissipation rates, can be approached by students in a first-year physics class by considering conservation laws,  and can therefore be useful for teaching about  energy transformations and conserved quantities. We then illustrate how to extend this method by applying it to a final example.

\end{abstract}

\maketitle

\section{Introduction}

Conservation of energy is a cornerstone of the sciences, and is therefore  a key concept in introductory physics classes. In practice, however, energy transformations can be difficult to track. It is therefore important to clearly distinguish which forms of energy are being considered in a problem and under what conditions they are constant for a given system.\cite{Bauman1, Bauman2,Chabay,Jewett, Seeley} For example, in sliding friction or inelastic collisions on a horizontal surface, the kinetic energy (KE) of the system is not constant; rather some of it is converted to underlying hidden degrees of freedom of the system and its surroundings. In this paper, we will discuss a class of systems in which  a constant of motion, frequently given by a conservation law (e.g. mass, charge, momentum, or angular momentum), provides a constraint that governs the exchange of energy between two subsystems as the system arrives at a new equilibrium. We will clearly designate the energy associated with the state of the system considered in each case by the forms of energy (either potential or kinetic) associated with the relatively easily measurable macroscopic properties of the system. We consider this energy to be ``dissipated'' when it is transferred to hidden internal degrees of freedom of the system and its surroundings in forms such as thermal energy. \cite{Tobin} We will see that, while energy is dissipated in each case, the total decrease of kinetic or potential energy in each process can be uniquely determined from the constant of the system.

We discuss five examples of this class of systems that can be introduced at the high school or introductory college physics level and use them to demonstrate a striking mathematical universality among these systems. These five systems and the mathematical formalism underlying them were previously noted and analyzed;\cite{bonanno2012} in this paper we discuss how pedagogical lessons from this suite of examples can be thematically integrated into an introductory physics curriculum and can be extended to lessons learned in later years and more advanced mathematical techniques.

We motivate this discussion by beginning with two closely related problems involving loss of KE during linear and rotational collisions in Sec.~\ref{sec:mechanics}, and discuss some pedagogical lessons that can be learned from each. We generalize these examples to identify a general mathematical formalism in Sec.~\ref{sec:general}. In Sec.~\ref{sec:hydroanalog}, we discuss two additional examples from fluid mechanics and electrical circuits that share these features  again highlighting lessons relevant to high school or undergraduate physics course work. These first four problems are introduced in the order that students may encounter them in a standard first-year physics sequence, thus showing how this class of examples could be used to provide a theme running through a first year curriculum.
Finally in Sec.~\ref{sec:springs}, we  demonstrate how to apply the formalism in general by considering a final example of two springs tied to a mass and immersed in a fluid, resulting in a damped oscillatory system. 

\section{Two Introductory Examples from Mechanics}\label{sec:mechanics}

We begin with two introductory mechanics problems, one dealing with translational motion and the other with rotational motion. These are textbook examples of first-semester introductory mechanics problems that we will use to identify key features that are universal to a broader class of physics systems.

\subsection{Loss of KE in Two Common Mechanics Problems}

\subsubsection{Loss of KE in a Perfectly Inelastic Collision Between Two Objects in One-Dimension}
\label{eX:PerfectlyInelasticCollision}

In a problem familiar to introductory physics students, we consider the example of a collision between two 
objects moving along a horizontal $x$-axis. Object $1$ has mass $m_1$ and is moving with an initial velocity $v_{1x}$ before collision, while
object $2$ has mass $m_2$ and is moving with an initial velocity $v_{2x}$ prior to collision with object $1$. The initial KE of the system in this reference frame is given by 
\begin{equation}
E_i=\frac{1}{2}m_1 {v^2_{1x}}+\frac{1}{2}m_2 {v^2_{2x}}.
\end{equation}

A perfectly inelastic collision has the additional constraint that the final velocities of the two objects must be equal. We assume an isolated system in which momentum is exchanged only between the two objects, so that the momentum of the two-object system is constant. If the objects collide head-on, their common final velocity is given by
\begin{equation}
v_{fx}=\frac{m_1 v_{1x}+m_2 v_{2x}}{m_1+m_2},
\label{eq:linearvelocity}
\end{equation} a result familiar to students in an introductory physics class.
The final KE is given by 
\begin{equation}
E_f=\frac{1}{2}\frac{m^2_1}{m_1+m_2}v^2_{1x}+\frac{1}{2}\frac{m^2_2}{m_1+m_2}v^2_{2x}+\frac{m_1 m_2}{m_1+m_2}v_{1x}v_{2x}~. 
\end{equation}
Therefore, the amount of KE dissipated in the collision is given by
\begin{equation}
\Delta E =E_f-E_i=-\, \frac{1}{2}\, \frac{m_1 m_2}{m_1+m_2} \, \left(v_{1x}-v_{2x}\right)^2.
\label{eq:collisiondissipation}\end{equation}

\subsubsection{Loss of KE in a System of Spinning Disks}
\label{ex:SpinningDisks}

For our second example, we consider an isolated system consisting of two rough disks lying parallel to each other, rotating about a common axis that goes through the center of each disk. Loss of KE in a similar system, in which a mass initially at rest is placed on a spinning disk, was examined in a recent paper.\cite{jain2012}
The first disk has moment of inertia $I_1$ and initial angular velocity $\omega_{1i}$, while the second disk has moment of inertia $I_2$ and initial angular velocity $\omega_{2i}$, so that the initial rotational KE of the disks is given by 
\begin{equation}
E_i=\frac{1}{2}I_1 {\omega^2_{1i}}+\frac{1}{2}I_2 {\omega^2_{2i}}.
\end{equation}
 
If the two disks are then put in frictional contact with each other  and isolated from their surroundings, their angular velocities will change until they reach equilibrium with a common angular velocity $\omega_f$, while  the angular momentum $I\omega$ of the two-disk system is constant, such that 
\begin{equation}
\omega_f=\frac{I_1 \omega_{1i}+I_2 \omega_{2i}}{I_1+I_2}.
\label{eq:linearvelocity}
\end{equation}
Since KE of the joint final state is given by $\frac{1}{2} \left(I_1+I_2\right) \omega_f^2$, the KE energy dissipated in this system can be calculated directly as  
\begin{equation}
\Delta E =-\frac{1}{2}\frac{I_1 I_2}{I_1+I_2}\left(\omega_{1i}-\omega_{2i}\right)^2.
\label{eq:collisiondissipationDisk}\end{equation}

Note that the decrease of KE given by Eq.~(\ref{eq:collisiondissipationDisk}) is the same form as that for the linear collision found in Eq.~(\ref{eq:collisiondissipation}), with the identification of $I$ with $m$, and of $\omega$ with $v$, as could be expected by a student with basic familiarity with rotational motion. 

\subsection{Methods of Dissipation in these Two Systems}

While the change of KE unsurprisingly takes the same form in each of these systems, the mechanism for this decrease may be wildly different. Notably, since the final state of  each isolated two-object system is constrained by conservation of momentum (angular or linear), the change in KE in these systems is independent of the mechanism. In this subsection, we explore some models of dissipation mechanism in these two systems.

\subsubsection{KE Dissipation in Collision}

The results obtained above were easily derived using algebraic methods and versions of this problem are common in introductory physics textbooks. The mechanism of this dissipation, however, can be very complicated.\cite{raman1920,goldsmith1960, pasha2014} A simplified model commonly used for inter-particle forces in partially inelastic granular materials is the linear spring-dashpot model,\cite{schwager2007} in which a linear spring produces a repelling elastic force, together with a dashpot that provides a damping force. However, the forces produced during deformation of colliding particles are not in general accurately modeled by linear relationship, so nonlinear relationships are sometimes used.\cite{hertz1882, renzo} 

The end state, however, is not sensitive to the model used. As an illustration, we assume that upon contact the particles are connected by a linear spring together with a damping force, given by 
\begin{equation}
F(\xi,\dot{\xi})=-k\xi-\gamma\dot{\xi},
\end{equation} 
where $\xi$ is the deformation of the spring  from its equilibrium length and the dots indicate time derivatives. This differs from the commonly used spring-dashpot model in that we allow the spring to supply both attracting and repelling forces. The spring constant $k$ and the damping constant $\gamma$ are taken to be properties of the material. With initial conditions, ${\xi}[0]=0$ and $\dot{\xi}[0]=v_{\rm rel}$, we find 
	\begin{equation}
	{\xi}[t] = - \frac{\mu v_{\rm rel} \,e^{- \frac{\gamma}{2\mu} \,t} }{\sqrt{\gamma^2-4 k \mu}}  \, \sinh\left( \frac{{\sqrt{\gamma^2-4 k \mu}}}{2\mu}\, t\right) ~,
	\end{equation} 
	where $v_{\rm rel}$ represents the relative approach velocity and $\mu\equiv\frac{m_1 m_2}{m_1+m_2}$ is the reduced mass of the system. The energy dissipated in the process does not depend on the particle parameters $k$ and $\gamma$, and is given by
	$$
	\int_0^\infty F(\xi)\,\dot{\xi}\, dt = \int_0^\infty \left( -k\xi-\gamma\dot{\xi} \right)  \dot{\xi}\, dt = -\frac12\, \mu\, v_{\rm rel}^2
	$$
	just as derived in Eq. (\ref{eq:collisiondissipation}).

\subsubsection{KE Dissipation in Spinning Disks}
While the above derivation above may be considered a relatively simply model for inelastic collisions compared to more realistic models, it still required solving differential equations well beyond the scope of most introductory physics classes. By contrast, for the rotational system, we consider a simplified dry friction model \cite{mungan} that generates a constant torque on the system and brings the disks to common angular velocity linearly in time. This case of constant torque can be analyzed using algebra as a case of rotational kinematics.  The initial angular velocities of disk $1$  and disk $2$ are $\omega_{1}$ and $\omega_{2}$, respectively, where we label the disks such that $\omega_{1} > \omega_{2}$. If there is a torque of constant magnitude $\tau$ acting between the disks, their angular accelerations will be of opposite signs with their magnitudes given by $\left|\alpha_1\right|=\tau/I_1$ and $\left|\alpha_2\right|=\tau/I_2$. The common angular velocity after time $\Delta t$ is given by $\omega_{f}=\omega_{1}-\tau\Delta t/I_1=\omega_{2}+\tau\Delta t / I_2$. The change in KE is $\Delta E =-\frac{1}{2}\frac{I_1 I_2}{I_1+I_2}\left(\omega_{1i}-\omega_{2i}\right)^2$, which is identical to Eq.~(\ref{eq:collisiondissipationDisk}).

\subsection{Some Lessons From Introductory Mechanical Examples}
A perfectly inelastic collision between two objects in one dimension is a problem that is commonly presented to students in both calculus- and algebra-based introductory physics classes at high school and college levels. The difficulty of deriving the reduction in KE  directly from the forces (requiring training in differential equations frequently not seen by physics majors until their second year, and not taken at all by many students who would take an algebra-based introductory physics class) can be contrasted with the simplicity of the approach based on conservation of momentum. Such a discussion with students can help to stress the power of conservation laws in greatly simplifying many problems.

The problem of the spinning disks, unlike the case of a linear collision, can be solved directly from the interaction model algebraically in a fairly realistic manner using a simplified friction model. This problem can thus demonstrate the agreement between the result obtained from momentum conservation and the result obtained from integrating forces directly from a specific dissipation rate in an example that students in a non-calculus based introductory course can solve for themselves.

Comparing these two problems can provide additional insight. Although the dissipation mechanisms and its time-dependent rate are very different, these differences in no way affect the total loss of KE, which is given by a common mathematical form. Consideration of other dissipation mechanisms, such as other models for the linear collision or velocity-dependent frictional models for the spinning disks, would similarly show that the specifics of the model do not affect the energy dissipated. Rather, the total loss is determined solely by the conservation of momentum (linear or angular) and the coupling between the two objects that resulted in a common final velocity (linear or angular). 

\section{General Formalism}\label{sec:general}

We noted earlier the similarity in form between the net change in KE for linear collision given in Eq.~(\ref{eq:collisiondissipation}) and that given in Eq.~(\ref{eq:collisiondissipationDisk}) for the spinning disks. 
These comparable results may not seem surprising, given the superficial similarity of these two systems. The comparison relied on the common substitution of rotational variables for their linear equivalents, such as the replacement of  ``momentum'' with ``angular momentum.'' Moving beyond this superficial similarity, we now generalize these results to identify the underlying commonality of a broader class of systems, of which these are but two examples. In doing so, we see that a variety of constants of motion can play a role equivalent to that of momentum in the linear collision and see that  this generalized ``momentum'' can lead to equivalent results for systems that at first glance seem strikingly dissimilar.

Generalizing from the collision examples, consider a system consisting of two subsystems that exchange energy with each other, and whose macroscopic states are fully described by two variables: ${\cal M}$ and ${\cal V}$. Here ${\cal M}_1$ and ${\cal M}_2$ are properties intrinsic to subsystems 1 and 2 respectively, while ${\cal V}_1$ and ${\cal V}_2$ may change during the interaction. The two subsystems begin with different initial values of ${\cal V}$, but then are brought into contact with each other until they reach a final equilibrium state that is defined by the common value of quantity ${\cal V}$, i.e., ${\cal V}_{1 f}={\cal V}_{2 f}={\cal V}_f$.

Since ${\cal M}_1$ and ${\cal M}_2$ are fixed, for given initial conditions ${\cal M}_1$, ${\cal M}_2$, ${\cal V}_{1i}$, ${\cal V}_{2i}$, there is only one degree of freedom in the final state: ${\cal V}_f$. Therefore, if some combination of ${\cal M}$ and ${\cal V}$ must remain constant in a given interaction, the change in kinetic or potential energy associated with the macroscopic state of the system will be uniquely determined by the initial values of the system and that constant quantity. Let us assume that we can choose variables ${\cal M}$ and ${\cal V}$ such that the quantity ${\cal M}{\cal V}={\cal M}_1{\cal V}_1+{\cal M}_2{\cal V}_2$ of the system as a whole remains constant and the kinetic or potential energy of each subsystem is given by $\frac 12 {\cal M}_k\,{\cal V}^2_k$, where $k=1,2$.
In this case, ${\cal M}_1{\cal V}_{1i}+ {\cal M}_2{\cal V}_{2i}= \left({\cal M}_1+{\cal M}_2\right){\cal V}_{f}$.  Once the system reaches equilibrium, the change in kinetic or potential energy of the system is given by
\begin{equation}
\Delta E  = \frac 12 {\cal M}_1\,{\cal V}^2_{1i} + \frac 12 {\cal M}_2\,{\cal V}^2_{2i}- \frac 12 \left({\cal M}_1 + {\cal M}_2\right)\,{\cal V}^2_f =
-\, \frac{1}{2}\, \frac{{\cal M}_1 {\cal M}_2}{{\cal M}_1+{\cal M}_2} \, {\cal V}_{rel}^2,
\label{eq:deltaE}\end{equation}
where ${\cal V}_{rel}\equiv{\cal V}_{1i} - {\cal V}_{2i}$.
Thus, independent of how the system comes to an equilibrium, the amount of energy dissipated during the process is completely determined.   

The two examples we have considered thus far fit the description above, with the identification of ${\cal M}$ as $m$ or $I$ and ${\cal V}$ as $v$ or $\omega$ for the linear or rotational collisions, respectively. We now show how this general formalism applies to other common problems in physics.

\section{Two Analogous Problems: One From Fluids and One From Electrical Circuits}\label{sec:hydroanalog}

Fluid flow is commonly used in introductory classes as an analogy to describe the flow of electrical current in a circuit.\cite{Serway, Walker} In this section, we look at two analogous systems: fluid flow between two reservoirs, and the electrical current between two capacitors. We see that both systems match our general formalism above, and that the similarity between these two systems yields additional insight into the process of dissipation of energy  in these systems.

\subsection{Application of the General Formalism}

\subsubsection{Application of the General Formalism to Fluid Flow Between Reservoirs} \label{ex:FluidReservoirs}
\begin{figure}[htb]
	\centering
	\includegraphics{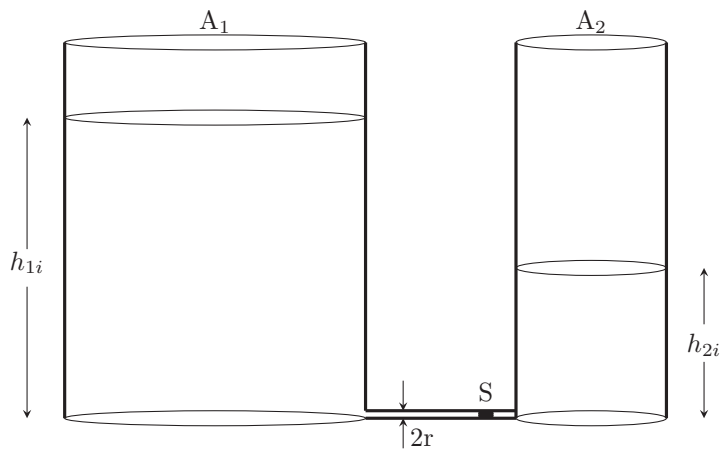}\\
	\caption{\label{tanksetup1}
				Tank 1, with cross-sectional area $A_1$,
				is filled to an initial height $h_{1i}$ with water.
				Tank 2, with cross-sectional area $A_2$ is
				initially filled to a height $h_{2i}$.  The tanks are connected with a pipe consisting of a
				circular cylinder with length $L$ and radius $r$, although a stopcock S
				initially prevents flow from tank 1 to tank 2.}
\end{figure}

\begin{figure}[htb]
	\centering
	\includegraphics{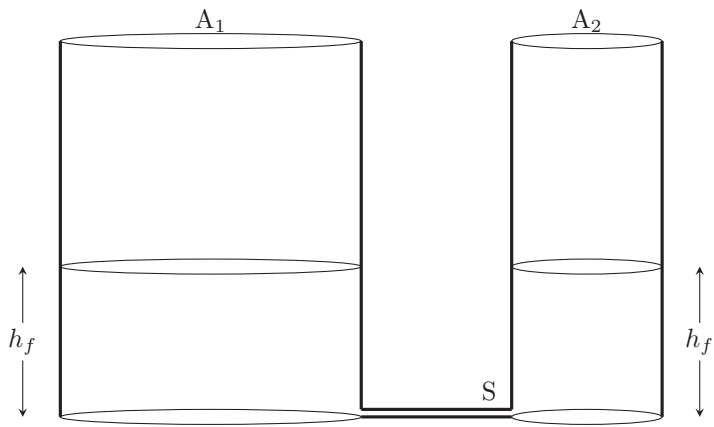}\\
	\caption{\label{tankfinal}
		Once the stopcock is opened, current will flow from tank 1
		to tank 2, until the system reaches the final state
		with both tank 1 and tank 2 filled to height $h_f$.}
\end{figure}

Consider a system consisting of tank $1$ with cross-sectional area $A_1$ and tank $2$ with
cross-sectional area $A_2$
joined by a long, narrow,
cylindrical pipe of length $L$ and radius $r$ in a uniform gravitational field of magnitude $g$. Tank 1
is initially filled with an incompressible fluid {such as water} of density $\rho$
to a height $h_{1i},$ and tank 2 to a lower height $h_{2i}$, as shown in Fig.~\ref{tanksetup1}.  A stopcock prevents the flow of water from tank 1 to 2. Once the stopcock is opened, water flows through the pipe, reaching the
final distribution in Fig.~\ref{tankfinal} with both tanks
filled to height $h_f$.

\newpage

Since no mass leaves the containers and $\rho$ and the acceleration of gravity $g$ are constant, conservation of mass gives us three equivalent choices for our constant quantity ${\cal M}{\cal V}$: the total  weight, mass, or volume of the water. These correspond to choices for our variables ${\cal M}$ and ${\cal V}$: in case 1,  ${\cal M}\equiv\rho g A$ and ${\cal V}\equiv h$; in case 2, ${\cal M}\equiv\rho A/g$ and ${\cal V}\equiv gh$; and in case 3, ${\cal M}\equiv A/(\rho g)$ and ${\cal V}\equiv \rho g h$.
Here ${\cal V}$ represents height of the water level $h$, gravitational potential at $h$, and hydrostatic pressure at the bottom of the reservoir, respectively; in each case ${\cal M}$ is intrinsic to each subsystem and ${\cal V}_{1f}={\cal V}_{2f}={\cal V}_f$.

In all three cases, the product ${\cal M} {\cal V}$ is constant for the entire system; in case 1 ${\cal M} {\cal V}=\rho g A h$ is the weight of the water, in case 2 ${\cal M} {\cal V}=\rho g h$ is the mass of the water, and in case 3 ${\cal M} {\cal V}=A h$ is the volume of the water.
In all three cases, the gravitational potential energy is given by $\frac{1}{2}{\cal M}{\cal V}^2=\frac{1}{2}\rho g A h^2$. 

\begin{figure}[htb]
	\centering
	\includegraphics{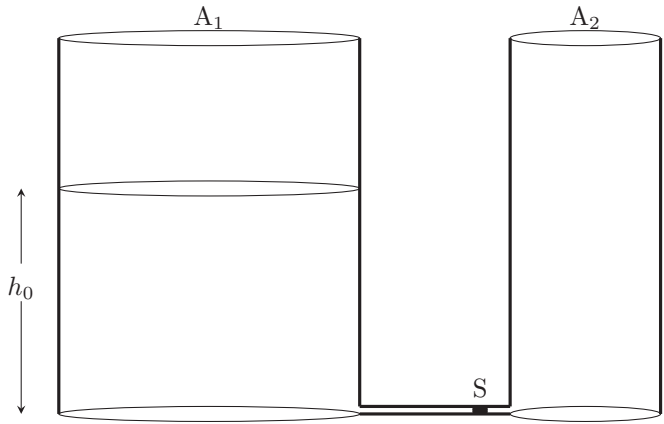}\\
	\caption{\label{tanksetup2}
		Tank 1 is filled to an initial height $h_0$ with water. Tank 2 is initially empty. }
\end{figure}

As the fluid is incompressible, ${\cal M}_1$ and ${\cal M}_2$ do not change, and as the fluid begins and ends at rest, KE at the start and end of the process is zero, so we only need to consider gravitational potential energy. Therefore this problem follows the formalism in Sec.~\ref{sec:general} and we can directly write the change in potential energy in any of our three cases as
\begin{equation}
\Delta E  = 
-\, \frac{1}{2}\,  
\frac{{\cal M}_1 {\cal M}_2}{{\cal M}_1+{\cal M}_2} \, {\cal V}_{rel}^2=-\frac{ \,\rho g\,}{2}\frac{A_1 A_2}{A_1+A_2}\,h_{rel}^2.
\label{eq:deltaEtanks}
\end{equation}
Without loss of generality, we can choose $h_{2i}=0$ and $h_{1i}=h_0$, as shown in Fig.~\ref{tanksetup2}, in which case $h_{rel}=h_0$; we adopt this choice for the remainder of this section. Comparison of Fig.~\ref{tanksetup2} with Fig.~\ref{tankfinal} makes it obvious that the potential energy of the system must have decreased in this process, since the height of the center of mass has visibly descended.

\subsubsection{Application of the General Formalism to Capacitor Discharge}\label{ex:CapacitorDischarge}
\begin{figure} [htb]
	\includegraphics[width=3in]{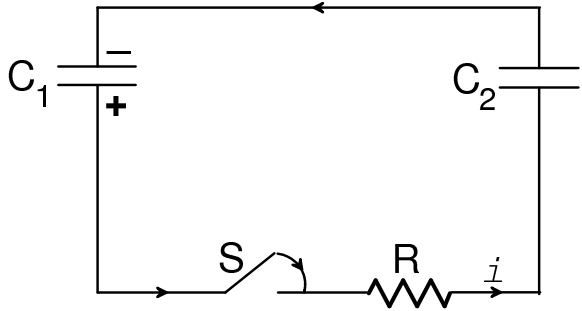}\\
	\caption{\label{capacitorcircuit}
		A circuit connecting two capacitors $C_1$ and $C_2$ in series
		with a resistor $R$.  Initially, $C_1$ holds charge $q_0$ and $C_2$ is
		uncharged.  The switch is initially open, so that no current flows. The switch is closed at time $t=0$, at which point current $i$ begins
		to flow.}
\end{figure}

As our fourth example, we analyze a circuit problem that is again 
suitable for presentation 
in an introductory course. Consider a circuit with two capacitors
$C_1$ and $C_2$ that are connected by a resistor
$R$ and a switch $S$. A schematic diagram is given in
Fig.~\ref{capacitorcircuit}. The capacitor $C_1$ is
initially charged to a voltage $V_{1i}$, and capacitor $C_2$ is initially 
charged to voltage $V_{2i}$. As in the fluid example, we can simplify the problem by
taking the case where capacitor $C_{1}$ is
initially charged to a value $q_{0} = C_1\, V_{0}$, where $V_{0}$ is
the initial voltage across $C_1$, and 
capacitor $C_{2}$ is initially uncharged.

When the switch is closed, a current $i$ ensues, and the circuit approaches an equilibrium with a common
potential difference $V_f$ across both capacitors. Since charge $q$ is conserved, we identify it with the ``momentum''  $q=C V = {\cal M}{\cal V}.$ Since capacitance is an unchanging geometric property of each capacitor, we identify 
${\cal M}\equiv C$ and therefore ${\cal V}\equiv V$. The electrostatic potential energy stored in a capacitor is given by 
$q=\frac{1}{2}C V^2 = \frac{1}{2}{\cal M}{\cal V}^2$. Therefore, the change of potential energy in this process is given by 

\begin{equation}
\Delta E  = 
-\, \frac{1}{2}\, \frac{{\cal M}_1 {\cal M}_2}{{\cal M}_1+{\cal M}_2} \, {\cal V}_{rel}^2=-\frac{1}{2}~\frac{C_1 C_2}{C_1+C_2}~V_0^2.
\label{eq:deltaEcircuit}
\end{equation}

\subsection{Energy Dissipation in Fluids and Electrical Circuits}

\subsubsection{Energy Dissipation in Fluid Flow Between Reservoirs}

We assume the flow to be quasi-static, and hence laminar everywhere; energy is  dissipated only  due to frictional flow through the pipe (frictional losses
in the tanks and minor losses at the inlet and outlet are negligible). With these assumptions, the pressure $p_1$ at the inlet of the pipe should be equal to the hydrostatic pressure $\rho g h_1$ and pressure at the outlet $p_2$ should be equal to $\rho g h_2$, where $h_1(t)$ and $h_2(t)$ are the
heights of the water level in tank 1 and tank 2, respectively.
In this case, 
\begin{equation}
\rho g h_1 + \frac12 \rho v_1^2  =
\rho g h_2 + \frac12 \rho v_2^2 + p_{loss},  \label{Energy-Loss-Pipe1}
\end{equation}
where $p_{loss}$ is the pressure loss due to frictional flow in the pipe,
and $v_1$ and $v_2$ are the spatially averaged instantaneous speeds of the fluid flow at the inlet
and outlet, respectively.

Let us denote the flow rate,
i.e., the volume of water that flows through the pipe per unit time,
by 
\begin{equation}
Q\equiv\pi r^2 v=-A_{1}\frac{dh_1}{dt}=A_{2}\frac{dh_2}{dt}.
\end{equation}
For laminar flow in a horizontal pipe of constant inner radius $r$, the pressure loss is given by Poiseuille's law: \cite{munson}
$p_{loss}=\frac{8 \mu L Q}{\pi r^4}$, or $p_{loss}=\alpha Q$, where
$\alpha\equiv\frac{8 \mu L}{\pi r^4}$.  
Since the fluid is incompressible, if $r$ is constant,
then the velocities $v_1$ and $v_2$ are equal and Eq. (\ref{Energy-Loss-Pipe1}) reduces to
\begin{eqnarray}
	\rho g h_1 =\rho g h_2 + \alpha Q =
	\rho g h_2 -\alpha A_{1}\frac{dh_{1}}{dt}.
	\label{Energy-Loss-Pipe2}
\end{eqnarray}

In a common analogy with electrical circuits, the reservoir plays the role of a capacitor, the pipe plays the role of a resistor, and the water plays the role of electrical charge. In our case 3, the quantity ${\cal M}\equiv A/(\rho g)$ is sometimes called ``hydraulic capacitance'' and the pressure ${\cal V}\equiv \rho g h$ plays the role of electrical potential.\cite{fuchs} With this choice, we can write Eq.~\ref{Energy-Loss-Pipe2} as
\begin{eqnarray}
{\cal V}_{1}={\cal V}_{2}-\alpha {\cal M}_{1}\frac{d {\cal V}_{1}}{dt}.
\label{pipeode}
\end{eqnarray}
Since ${\cal M}{\cal V}$ is conserved, we can write ${\cal V}_{2}=\frac{{\cal M}_{1}}{{\cal M}_{2}}\left({\cal V}_{0}-{\cal V}_{1}\right),$ where ${\cal V}_{0}\equiv{\cal V}_{1i}$. Solving this differential equation, we get
$${\cal V}_1 = {\cal V}_0  \left(\frac{{\cal M}_1+{\cal M}_2\, e^{-\frac{t}{\tau}}}{{\cal M}_1+{\cal M}_2} \right) ~\mbox{and}~
{\cal V}_2 = {\cal V}_0  \left(\frac{{\cal M}_1-{\cal M}_1\, e^{-\frac{t}{\tau}}}{{\cal M}_1+{\cal M}_2} \right),  $$
 where
$\tau \equiv \alpha\left(\frac{{\cal M}_1{\cal M}_2}{{\cal M}_1+{\cal M}_2}\right).$
The flow rate is then given by
\begin{eqnarray}
Q = \frac{1}{\alpha} {\cal V}_0 e^{-\frac{t}{\tau}}.
\label{flowrate}
\end{eqnarray}
The net power applied on the fluid inside the pipe  is given by the product of $p_{loss}$ and the flow rate $Q$, and the total gravitational potential energy dissipated in the entire transfer process is then given by
\begin{eqnarray}
\Delta E &=& -\int_0^\infty Q^2 \; \alpha \; dt
\end{eqnarray}
which simplifies to
\begin{equation}
\Delta E  = 
-\, \frac{1}{2}\,  
\frac{{\cal M}_1 {\cal M}_2}{{\cal M}_1+{\cal M}_2} \, {\cal V}_{0}^2=-\frac{ \,\rho g\,}{2}\frac{A_1 A_2}{A_1+A_2}h_{rel}^2.
\label{Energy-Loss-Hydro}
\end{equation}

We again see that the frictional losses in the pipe are equal to the decrease in potential energy found in Eq. (\ref{eq:deltaEtanks}) and are independent of $\alpha$, a measure of the restrictive property of the pipe.

\subsubsection{Energy Dissipation in capacitor discharge}

Note that unlike several of the previous problems, the instantaneous rate of power transferred to thermal energy in the resistor $P=i^2 R$ is well known to students in introductory physics classes.
The instantaneous power loss depends on the resistance both directly and through the current $i$; however, as we have seen, the total potential energy dissipated is unaffected by the details of the dissipation rate.
We now
show that the results we obtained in Eq. (\ref{eq:deltaEcircuit}) are
consistent with the results obtained by integrating the
power dissipated  over time.

As the capacitor discharges, we characterize the system at
an intermediate time $t$, where the potential difference across capacitors $C_1$ and $C_2$ are
$V_1\left(t\right)$ and $V_2\left(t\right)$ respectively, and there is a current
$i\left(t\right)$ through the resistor.
Kirchoff's loop rule yields
\begin{eqnarray}
V_1  = i\, R + V_2 ~. \nonumber
\label{Kirchoff}
\end{eqnarray}
Using $i\left(t\right) = -\frac{dq_1}{dt}=-C_1\frac{dV_1}{dt}$,
and, from charge conservation,
 $C_1 V_1+C_2 V_2=C_1 V_0$, we get 
\begin{equation}
V_{1}=V_{2}-R C_{1}\frac{d V_{1}}{dt}.
\label{circuitode}
\end{equation}
With the identification of electric potential ${\cal V} \equiv V$, and capacitance ${\cal M}\equiv C$ as our variables, and with the resistance $R$ playing the role that $\alpha$ played for pipe flow, this is equivalent to the differential equation given in Eq.~(\ref{pipeode}).

Assuming the discharge
is done in a quasi-static process to avoid
radiation, \cite{boykin2002, mita1999, gangopadhyaya2000, choy} 
and thus the only source of power dissipation is the resistor,  $P=i^2R$, the electrical analog to $Q^2\,\alpha$ for the fluid case. Following the steps given in Eqs.~(\ref{pipeode}-\ref{Energy-Loss-Hydro})  therefore demonstrates that the total reduction of electrostatic potential  energy in the discharge is
\begin{eqnarray}
\Delta E  = 
-\, \frac{1}{2}\,  
\frac{{\cal M}_1 {\cal M}_2}{{\cal M}_1+{\cal M}_2} \, {\cal V}_{0}^2=-\frac{1}{2}~\frac{C_1 C_2}{C_1+C_2}~V_0^2 .\label{Energy-Loss-EM}
\end{eqnarray}
This is equivalent to Eq. (\ref{eq:deltaEcircuit}) found previously,
indicating that the energy transferred to thermal energy in the resistor is
equal to the potential energy lost by the capacitors and is independent of $R$.

We once again see that the energy dissipation is determined solely by the initial state of the system, the coupling between the subsystems, and a constant quantity (in this case, electric charge). However, unlike the collision or fluid-flow cases, in which students will likely not be familiar with relevant dissipation models, the power dissipated in a resistor ($P=i^2R$) is a common element introduced in introductory classes. Since the instantaneous power dissipation does depend on resistance, this example highlights the power of this method in showing the total energy dissipated is independent of resistance.

\subsubsection{Some Lessons from Analogous Fluid and Electrical Flow}

One remarkable property of the fluid flow example is that the necessity of loss of gravitational potential energy can be easily visualized.  Visual comparison of Fig.~\ref{tanksetup2} with Fig.~\ref{tankfinal} shows immediately that the center-of-mass height of the fluid decreases in this process; hence the gravitational potential energy must decrease.  Moreover, the magnitude of the loss is determined directly by the initial and final states of the system. This visual illustration of the principle can be very useful for helping introductory students understand how loss is determined in these systems, and can shed light on the entire suite discussed in this paper. 

Furthermore, many students may not be familiar with the process of power dissipation in a frictional flow through a pipe.  While students in introductory classes will not generally be familiar with techniques to solve differential equations, students in calculus-based physics classes can test that the differential equation is solved by the given solution that satisfies the initial conditions. For more advanced students, then, this problem can serve as a useful introduction to frictional losses in a pipe.  

This example is particularly instructive when combined with the example of the coupled capacitors. In a common analogy, fluid flow through a pipe can be compared to the flow of electrical current through a wire; here the volume current $Q$ plays the role of electrical current $i$, with ``hydraulic capacitance'' substituting for $C$ and pressure for $V$. In this case, the functional form of not only the potential energy, but also of the power dissipated and differential equations solved are directly analogous, as can be seen by comparing Eq.~(\ref{pipeode}) to Eq.~(\ref{circuitode}). The rate of energy dissipation in the frictional fluid flow $Q^2 \alpha$ compares directly to that in the circuit, $i^2 R$, which is familiar to students in introductory physics classes.  
Here $\alpha$ contains all of the information about the size 
and frictional properties of the pipe flow, and plays a role similar to that of resistance $R$ in the
electrical circuit.

This analogy can help build intuition regarding both systems. The ability to visualize the loss potential energy by loss of center-of-mass height can assist in developing an understanding of the role of dissipation in the capacitor example, where it is not so easily visualized. Additionally, the familiarity of students with power dissipation in a resistor can help students to gain insight into the less-familiar loss of power in fluid flow through a pipe.

\section{Extension to the method: A Two-Spring System}\label{sec:springs}
\begin{figure}[htb]
	\centering
	\includegraphics{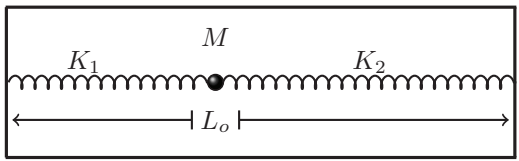}\\
	\caption{Two springs with force constants $K_1$ and $K_2$ are connected to a mass $M$ with fixed rigid supports on either side. The motion takes place inside a box filled with a fluid.}
	\label{fig.springs}
\end{figure}

Thus far, we have considered four systems, three of which show up in  introductory textbooks, and the loss of gravitational potential energy in  the remaining case (the fluid system) is very easy to visualize. The surprising universality of this category of systems can also extend beyond these textbook examples. We now consider a system that consists of two  Hooke's Law springs connected to a mass, with the other ends of the springs tied to rigid walls, as shown in Fig. \ref{fig.springs};
The entire system is immersed in a fluid bath that generates a velocity-dependent frictional force.  
We will apply the general formalism of Sec. \ref{sec:general} to identify the state variables  ${\cal M}$ and ${\cal V}$ for this system, and then determine the potential energy change for the system as it moves from a state of higher potential energy to equilibrium. 

Let the unstretched lengths of the springs be $L_1$ and $L_2$. If the springs are initially stretched by  $x_{1i}$ and $x_{2i}$ respectively, we must have $L_1+x_{1i}+L_2+x_{2i}=L_o$, where $L_o$ is the distance between the two vertical walls. When we release the system from the initial state, $x_1$ and $x_2$ change, but the sum $x_1+x_2 = L_o-L_1-L_2$ remains constant. This makes, $x_1$ and $x_2$ good candidates for quantities  ${\cal M}_1{\cal V}_1$ and ${\cal M}_2{\cal V}_2$ respectively. Comparing the potential energy of the springs $\frac12 K_jx_j^2$ with $\frac12 \, {\cal M}_j{\cal V}_j^2$, we are able to identify both ${\cal M}_j$ and ${\cal V}_j$:
\begin{equation}
{\cal V}_j = K_jx_j \quad \quad \mbox{and} \quad\quad {\cal M}_j = \frac{1}{K_j}~~,
\end{equation}
for $j=1,2$, where $K_j$ represents the stiffness constant of the $j$-th spring. 
Note that ${\cal M}_j=1/K_j$ is  intrinsic to each spring and the condition of equilibrium is ${\cal V}_{1f} = {\cal V}_{2f}$, as one would expect from Newton's second law.  This then immediately allows us to write down the change of potential energy to be 
\begin{equation}
\Delta E = -\, \frac{1}{2}\, \frac{{\cal M}_1 {\cal M}_2}{{\cal M}_1+{\cal M}_2} \, {\cal V}_{rel}^2\quad = \quad
-\frac12\,\frac{1}{K_1+K_2}~\left[ K_1 x_{1i}- K_2 x_{2i}\right]^2  ~~. \label{DeltaE1}
\end{equation}

 As an exercise, this result can be verified directly from the fact that the total length of the system is constant such that \begin{equation}x_{1i}+ x_{2i} = x_{1f}+ x_{2f}~~,\label{eq.spring1}\end{equation} 
and that forces balance at equilibrium so that 
\begin{equation}K_1 x_{1f} =  K_2 x_{2f}~~.\label{eq.spring2}\end{equation}

This example illustrates how to extend the general formalism beyond the four well-known examples demonstrated in the previous section. Discussing this process with students after having worked through the previous four examples throughout their introductory course sequence can help them to understand ways in which we can identify universal features in diverse systems.
   
\section{Conclusions}\label{sec:conclusions}

When subsystems interact via dissipative forces to reach a common equilibrium, the requirement of a constant of motion can determine the energy dissipated during the interaction process.  We have analyzed five examples of this type of system that share a common mathematical formalism, as shown in Table~\ref{table:summary}. The first four examples, or their variants, can be found in introductory textbooks. In each case, the systems are characterized by two state variables ${\cal M}$ and ${\cal V}$ such that the energy is given by $\frac 12 {\cal M}\,{\cal V}^2$, and the product ${\cal M}{\cal V}$ is constant throughout the interaction. In these cases, the change in the relevant form of energy, kinetic or potential, of the process was given by
\begin{equation}
\Delta E  =-\, \frac{1}{2}\, \frac{{\cal M}_1 {\cal M}_2}{{\cal M}_1+{\cal M}_2} \, {\cal V}_{rel}^2,
\nonumber\end{equation}
where ${\cal V}_{rel}={\cal V}_{1i}-{\cal V}_{2i}$.

\begin{table}[h!]
	\centering
\caption{The five systems discussed, relevant variables, and constants of motion.}
	\begin{ruledtabular}
		\begin{tabular}{l  l  l  l}
	~~~~~~~System & ~~~~~~~~~~~~~~~~ ${{\cal M}}$ &~~~~~~~~~ ${\cal V}$ & Constant Quantity ${\cal M}{\cal V}$ \\
	\hline
	Inelastic collision & Mass $m$ & Velocity $v$ & Momentum \\
	Co-rotating disks	& Moment of Inertia $I$ & Angular Velocity $\omega$ & Angular Momentum \\
	Fluid reservoirs  &&&\\
	\hspace{.5cm}$\boldsymbol{\cdot}$Case 1	& $\rho g A$ & height $h$ & Weight\\
	\hspace{.5cm}$\boldsymbol{\cdot}$Case 2	& $\rho A/ g$			& Grav. Potential $gh$ & Mass\\
	\hspace{.5cm}$\boldsymbol{\cdot}$Case 3	& $A/\rho g$			& Pressure $\rho gh$ & Volume\\
	Capacitors charging & Capacitance $C$ & Electric Potential $V$ & Charge \\
	Two-Springs System & Inverse of spring constant  $\frac1{K_j}$ & Spring force $K_jx_j$ & Displacement $x_j$ \\
		\end{tabular}
	\end{ruledtabular}
	\label{table:summary}
\end{table}

 This formalism could be introduced in the first semester of a physics sequence in the context of momentum and angular momentum, and then returned to when students encounter fluids and capacitors later in the curriculum, providing a theme throughout the first-year curriculum.
We used these very different examples to highlight useful pedagogical lessons, and extended our method to a fifth example that illustrates the process of identification of $\cal M$ and $\cal V$ for a given physical problem. Although the quantities represented by $\cal M$ and $\cal V$ were different for each specific system, the relevant form of energy is given by $1/2 {\cal M}{\cal V}^2$ in all cases.

This general result could be expanded beyond those discussed here. For instance, the case of an inelastic collision between two particles could be extended to a molecular level to discuss two interacting atoms coming together to form a diatomic molecule; in this case this formalism would indicate the binding energy of the molecule. 
The methodology discussed in this paper is therefore useful in analyzing a wide variety of different problems and can provide additional insight into various systems.

\begin{acknowledgments}
We would like to thank Constantin Rasinariu for carefully reading the manuscript and making very valuable comments. We also thank the referees for many insightful comments from which this manuscript has benefited greatly.

This work is an extension of an unpublished paper, in which we investigated the analogy between the hydrodynamic and two-capacitor systems in detail. \cite{unpublished}
 During the later stages of this work, we came across a paper \cite{bonanno2012} that analyzes the same five examples that we described here. We believe our work, with a focus on how these analogous systems can be used in an introductory physics curriculum, will be of additional value to students and faculty members.

\end{acknowledgments}


\end{document}